# Hybrid GRU-CNN Bilinear Parameters Initialization for Quantum Approximate Optimization Algorithm

Zuyu Xu, Pengnian Cai, Kang Sheng, Tao Yang, Yuanming Hu, Yunlai Zhu, Zuheng Wu, Yuehua Dai, Fei Yang*
School of Integrated Circuits, Anhui University, Hefei, Anhui, 230601, China.
E-mail: feiyang-0551@163.com

**Abstract**.

The Quantum Approximate Optimization Algorithm (QAOA), a pivotal paradigm in the realm of variational quantum algorithms (VQAs), offers promising computational advantages for tackling combinatorial optimization problems. Well-defined initial circuit parameters, responsible for preparing a parameterized quantum state encoding the solution, play a key role in optimizing QAOA. However, classical optimization techniques encounter challenges in discerning optimal parameters that align with the optimal solution. In this work, we propose a hybrid optimization approach that integrates Gated Recurrent Units (GRU), Convolutional Neural Networks (CNN), and a bilinear strategy as an innovative alternative to conventional optimizers for predicting optimal parameters of QAOA circuits. GRU serves to stochastically initialize favorable parameters for depth-1 circuits, while CNN predicts initial parameters for depth-2 circuits based on the optimized parameters of depth-1 circuits. To assess the efficacy of our approach, we conducted a comparative analysis with traditional initialization methods using QAOA on Erdős-Rényi graph instances, revealing superior optimal approximation ratios. We employ the bilinear strategy to initialize QAOA circuit parameters at greater depths, with reference parameters obtained from GRU-CNN optimization. This approach allows us to forecast parameters for a depth-12 QAOA circuit, yielding a remarkable approximation ratio of 0.998 across 10 qubits, which surpasses that of the random initialization strategy and the $PPN^2$ method at a depth of 10. The proposed hybrid GRU-CNN bilinear optimization method significantly improves the effectiveness and accuracy of parameters initialization, offering a promising iterative framework for QAOA that elevates its performance.

## 1. Introduction

With the advent of Noisy Intermediate-Scale Quantum (NISQ) devices, characterized by a limited number of noisy qubits capable of supporting only shallow-depth circuits, the demand for algorithms suitable for near-term quantum computing has surged. Variational Quantum Algorithms (VQAs) have emerged as a promising class of algorithms to harness these devices' capabilities and limitations[1, 2]. VQAs harness quantum processors to prepare quantum states defined by variational parameters, subsequently optimizing these parameters through classical computers in a closed-loop manner[3]. Among VQAs, the Quantum Approximate Optimization Algorithm (QAOA) has risen as a promising method to expedite the resolution of combinatorial optimization problems by providing approximate solutions[4]. QAOA has found applications across diverse optimization problems, including the traveling salesman problem, Max-Cut problem, portfolio optimization, among others[5-7]. Despite its promise as a quantum algorithm, practical implementations of QAOA grapple with challenges tied to the existing constraints of quantum hardware and algorithm optimization [8-10]. Foremost among these challenges is the task of determining appropriate initial parameters, capable of significantly enhancing both the convergence rate and accuracy of the approximation [11-13].

Recent research on the QAOA has garnered widespread attention for parameter optimization on NISQ devices[11, 14-17]. Hybrid approaches that meld quantum and classical computational resources, as well as introducing innovative algorithms, are being explored to bolster the practicality and efficacy of QAOA[3, 18-22]. Traditional optimization techniques, such as gradient-descent-based algorithm like Adam and Adagrad [23-25], have been conventionally employed for initializing QAOA parameters [26]. However, their efficiency is hampered by the demand for extensive circuit depths and measurements to compute derivatives of the output from variational quantum circuits [27]. In recent developments, the optimization of parameters for VQAs has increasingly leveraged neural networks, proving to be a potent tool for enhancing performance [11, 13, 15, 27, 28]. In Ref. [15], the Convolutional Neural Networks (CNNs) have been exploited for predicting QAOA parameters, which showcase its efficiency and advantages in comparison to existing classical optimizers. Nonetheless, it's noteworthy that QAOA parameters inherently constitute sequential data, a domain where Recurrent Neural Networks (RNNs) have demonstrated remarkable suitability for the optimization process within QAOA[13]. Multiple investigations have underscored the substantial reduction in the total number of optimization iterations required to achieve a specific level of accuracy when implementing RNN initializations, such as Long Short-Term Memory (LSTM) or Gated Recurrent Unit (GRU) configurations[13, 27]. However, as reported in [13], RNNs need to invoke quantum resources for each iteration when implementing parameter prediction, potentially leading to inefficiencies, especially when delving into higher circuit layers and thereby augmenting the training workload on the network.

More recently, X. Lee introduced a novel bilinear strategy for initializing deep QAOA circuit parameters [3]. Diverging from approaches that entail multiple trials to secure optimization, this strategy necessitates just a single trial at each depth, manifesting its superiority in terms of approximation and optimization costs. Nonetheless, this technique relies on utilizing the previous optimal parameters as a starting point for subsequent depths, resulting in limitations related to generalization and additional training costs. Hence, there exists a demand for more versatile and efficient initialization and optimization strategies within the domain of QAOA, which hold the promise of significantly expediting the pursuit of optimal solutions to intricate problems.

In this work, we introduce a novel hybrid optimization technique named the GRU-CNN bilinear strategy, applied to the initialization of QAOA parameters. This model adeptly optimizes the variational parameters of QAOA at depths 1 and 2, eliminating the need for an exhaustive search for precise initial parameters within the parameter space, particularly for Max-Cut problem instances. Subsequently, we conduct an exhaustive benchmark of our approach, demonstrating its superior performance compared to classical optimizers. We employ the initial variational parameters obtained from the GRU-CNN model at depths 1 and 2 as starting points for subsequent depths, which are utilized to predict the initial parameters of circuits at higher depths by bilinear algorithm. This prediction takes into account the observed pattern of variation in the optimal parameters, considering both the angle index (j) and the circuit's depth (l). The combined application of these techniques effectively initializes circuits at higher depths, preventing convergence to undesired local optimal values and thereby improving the overall performance of our approach. As a result, we evaluate the performance of QAOA on instances with depths ranging from 3 to 12, achieving approximate ratios up to 0.998. We compare the results obtained by our strategy with those of other optimization approaches on instances of depth 10. These outcomes underscore the promise of our initialization strategy as a highly effective means of addressing the Max-Cut problem within the QAOA framework. This work not only provides valuable insights into QAOA parameter initialization but also highlights the significant potential of our proposed strategy in addressing the Max-Cut problem.

The subsequent sections of this paper are structured as follows: Section 2 provides an overview of the Quantum Approximate Optimization Algorithm (QAOA) and its application to the Max-Cut problem. Section 3 outlines the novel GRU-CNN initialization optimization strategy, elucidating the network model and the associated loss function within this hybrid network architecture. Section 4 undertakes comprehensive testing of our proposed network preprocessing model using a diverse array of illustrative examples. Section 5 employs this algorithm to evaluate the performance of QAOA at increased depths, utilizing the parameter starting points generated via the GRU-CNN initialization technique and it was compared with other initialization strategies. Finally, Section 6 encapsulates our work and offers a glimpse into future directions. This reorganization aims to enhance the clarity and coherence of the paper's structure, aligning it with academic conventions.

## 2. QAOA for solving Max-Cut

2.1 The Quantum Approximate Optimization Algorithm

The Quantum Approximate Optimization Algorithm (QAOA) and its variants, which amalgamate the benefits of quantum annealing [22] and variational quantum algorithms, have been proposed for addressing combinatorial problems on Noisy Intermediate-Scale Quantum (NISQ) machines, holding considerable potential. Fig. 1 visualize the L-level QAOA. When applied to the Max-Cut problem, a QAOA of depth $L$ comprises $L$ layers of alternating operators in the initial state $|\varphi_0\rangle$:

$$|\varphi(\gamma,\beta)\rangle = (\prod_{l=1}^{L} e^{-i\beta_l H_M} e^{-i\gamma_l H_C})|+\rangle^{\otimes N} \quad (1)$$

where $|+\rangle^{\otimes N}$ is the result produced by the Hadmand gate of the initial state $|\varphi_0\rangle$ acting on $N$ qubits, $H_M = \sum_{n=1}^{N} \sigma_n^X$ is the mixing Hamiltonian, $H_C$ is the problem Hamiltonian, and $\beta_l$, $\gamma_l$ is the variational parameter of the quantum gate of the corresponding layer of the QAOA, the operator $e^{-i\beta_l H_M}$ can be realized by $R_X(\beta_l)$ acting on all the qubits at the same time. To implement the operator $e^{-i\gamma_l H_C}$, we follow this procedure:

$$e^{-i\gamma_l H_C} = \prod_{(i,j)\in E} \exp(-i\gamma_l(\frac{I - Z_i Z_j}{2})) \quad (2)$$

Thus, $e^{-i\gamma_l H_C}$ can be equated to $\text{CNOT}_{ij}(I_i \otimes R_Z(2\gamma_l)_j)\text{CNOT}_{ij}$, i.e., it can be realized with two *CNOT* entanglement gates and one *RZ*-gate[29].

To find the optimal solution, a classical optimizer is used to update $\gamma$ and $\beta$ by maximizing the objective function:

$$E_L(\gamma,\beta) = \langle \varphi_L(\gamma,\beta) | H_C | \varphi_L(\gamma,\beta) \rangle \quad (3)$$

In the optimal setting, $|\varphi(\gamma^*,\beta^*)\rangle = \arg\max E_L(\gamma,\beta)$.

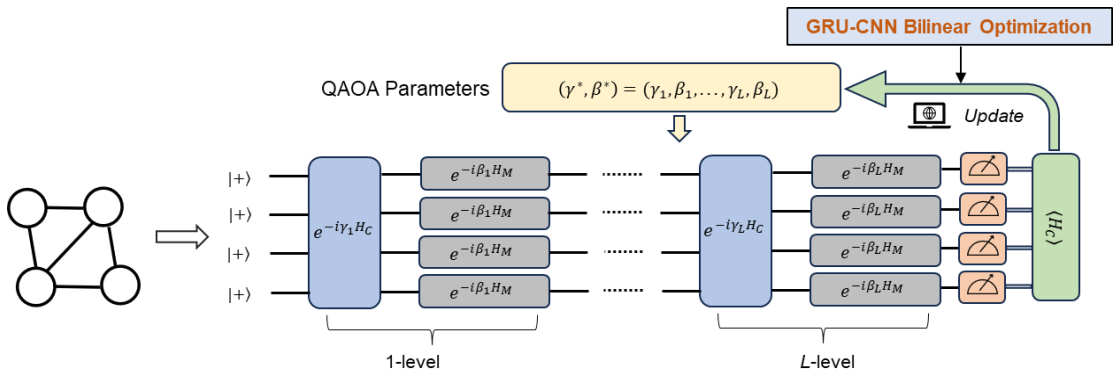

Figure 1: Schematic of a *L*-level QAOA for solving the Max-Cut problem. For a *L*-level QAOA, a quantum circuit operates on the input state $|+\rangle^{\otimes N}$, the $e^{-i\gamma_l H_C}$ and $e^{-i\beta_l H_M}$ are applied alternately to the initial superposition state, and measure the final

state to determine the expected value $\langle H_C \rangle$ with respect to the objective function. The GRU-CNN bilinear strategy uses the measured bitstring to update parameters of the circuit.

2.2 The Max-Cut Problem

The Max-Cut problem is one of the typical combinatorial optimization problems[30]. The problem aims to find the maximum cut of a (weighted or unweighted) graph, i.e., given a graph $G=(V,E)$, where $V$ is the set of vertices and $E$ is the set of edges, The (unweighted) maximum cut objective is the partition of the nodes into two groups [+1,-1] such that there are as many edges as possible between nodes from these two sets. The optimal solution $z^*$ of Max-Cut maximizes the objective function $C(z)$:

$$C(z) = \frac{1}{2} \sum_{(i,j) \in E} w_{ij}(1 - z_i z_j) \tag{4}$$

where $E$ is the set of edges of the given Erdős–Rényi graph and $w_{ij}$ is the edge weights.

The technical solution adopted by QAOA to solve the Max-Cut problem is to use the Pauli-$Z$ operator to construct the problem Hamiltonian quantity describing the Max-Cut problem is given by:

$$H_C = \frac{1}{2} \sum_{(i,j) \in E} w_{ij}(I - \sigma_i^z \sigma_j^z) \tag{5}$$

where $\sigma_i^Z$ refers to the Pauli-$Z$ operator applied on the ith qubit with $i \in [N]$. In this paper, we simply set the weight of all edges to be 1, i.e., $w_{ij}=1$.

2.3 Benchmarking

The common performance metric of QAOA for the Max-Cut problem is the approximation ratio $R$:

$$R = \frac{E_L(\gamma, \beta)}{C_{max}} \tag{6}$$

where $C_{max}$ is the maximum cut of the graph.

## 3. Using GRU-CNN to Predict the Initial Variational Parameters

In this section, we describe the architecture of the proposed hybrid GRU-CNN neural network model and the loss function.

3.1 Network Architecture

The goal of GRU-CNN is to optimize the QAOA parameters in depth-1 and depth-2. Our model incorporates two primary components: GRU and CNN, with the comprehensive architecture illustrated in Fig. 2.

To predict initial variational parameters for QAOA at depth 1, we utilize a GRU neural network [31], a variant of the recurrent neural network (RNN). The GRU is particularly adept at addressing issues related to vanishing and exploding gradients that can be problematic when training RNNs through backpropagation [32, 33]. A recurrent neural network, characterized by its feedback loops, excels at encoding contextual information from sequential data [34]. The equations expressing the GRU cell are as follows:

$$z_t = \sigma(W_z x_0 + R_z h_{t-1} + b_z + d_z) \tag{7}$$

$$r_t = \sigma(W_r x_o + R_r h_{t-1} + b_r + d_r) \tag{8}$$

$$\tilde{h}_t = \tanh(W_h x_0 + r_t \odot (R_h h_{t-1} + d_h) + b_h) \tag{9}$$

$$h_t = z_t \odot h_{t-1} + (1 - z_t) \odot \tilde{h}_t \tag{10}$$

where the hidden state $h_t$, and the input $x_o$ (in our case, the optimizable parameters and cost) are also shown. The symbols $\odot$, $\sigma$, and tanh denote the Hadamard product, the sigma function, and the hyperbolic tangent function, respectively. Initially, we introduce a bias vector $d$ to modulate the computation of $\tilde{h}_t$, which serves as a candidate state gate used to generate a candidate for a new hidden state. The update gate $z_t$ selectively retrieves information from the preceding hidden state. Likewise, the reset gate $r_t$ determines the information to be omitted from previous hidden states. The reset gate effectively eliminates less crucial information from previous iterations. In the final step, to compute the new hidden state $h_t$, we consider the update gate's role in determining the weighting between the candidate state $\tilde{h}_t$ and the prior hidden state $h_{t-1}$. Here, we utilize the GRU to stochastically generate an initial parameter set. We iterate through a predefined number of time steps to refine the initial parameters for QAOA at depth 1 ($l$=1). This iterative approach enables us to obtain high-quality initial parameters suitable as a starting point for further optimization. By employing the GRU and this iterative process, our objective is to enhance the quality of initial parameters and subsequently improve the performance of QAOA.

The network takes stochastically predicted initial variational parameters and the associated problem instance costs as inputs. It optimizes these initialized parameters for the specific instance across a defined time step, ultimately producing the optimized parameters $(\gamma_1, \beta_1)$. In each GRU iteration, the network receives as input the expectation of the cost function $E_t$, computed by the preceding QAOA, where $E_t$ estimates $\langle H_C \rangle$, and the parameter $\theta_t$ used in the QAOA evaluation. The GRU unit also leverages information stored in its internal hidden state from the previous time step $h_t$, which contains trainable parameters, facilitating the parameter mapping process as:

$$[\theta_{t+1}, h_{t+1}] = GRU(\theta_t, h_t, E_t) \tag{11}$$

As shown in Fig. 2, the GRU generates optimized parameters for the initial parameters (p=1) of the QAOA, along with a new hidden state. After proposing this new set of QAOA parameters, the GRU forwards them to the QAOA for evaluation, and the iterative loop progresses.

Initial variational parameters for QAOA at depth 2 are predicted using a convolutional neural network (CNN) [35]. Given that QAOA parameters fall within a continuous interval and exhibit interdependencies, CNNs can effectively extract pertinent features[36, 37]. Our CNN architecture comprises two primary components. The first part employs two convolutional layers to extract features from the input QAOA parameters at depth 1. The initial layer expands the input dimensions from 1 to 16, followed by the second layer, which further extends dimensions to 64. Each layer is equipped with 2x2 kernels. In the second part of our architecture, 3x2 kernels are applied to reduce dimensionality from 64 to 1, yielding the final network output. Notably, all convolutional layers in our design feature a step size of 1. Additionally, we utilize the Rectified Linear Unit (ReLU) activation function for nonlinear mapping [38].

The hybrid architecture proposed in this paper capitalizes on the complementary strengths of GRU and CNN networks to enhance the optimization of initial QAOA circuit parameters. As a preprocessing strategy, the GRU-CNN architecture harnesses the GRU's ability to effectively converge with stochastic initialization. Specifically, at depth 1, the GRU optimizes the prediction of initial QAOA parameters. The well-optimized parameters derived from the GRU model provide a robust starting point for predicting parameters at higher depths. Subsequently, these parameters are input into the CNN, which excels at forecasting parameters for QAOA at depth 2 and is adept at handling intricate patterns and dependencies. The iterative procedure, beginning with parameter optimization at depth 1 using the GRU and then employing these optimized parameters for predicting parameters at depth 2 with the CNN, facilitates efficient and precise parameter estimation for the QAOA circuit. These optimized parameters from GRU-CNN serve as the starting point for the bilinear algorithm, which will be discussed in Section 6.

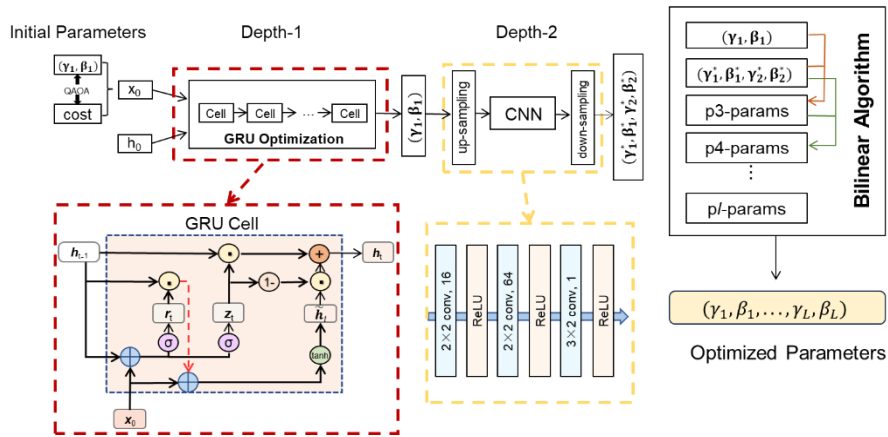

Figure 2: This diagram illustrates our optimization strategy, outlining the process of predicting and optimizing initial parameters for circuits at depths 1 and 2. It employs a hybrid network structure that combines a gated recurrent unit (GRU) and a convolutional neural network (CNN). The diagram also provides a schematic of our GRU unit and an overview of the internal network structure of the CNN. Moreover, it demonstrates the iterative prediction of optimized parameters for circuits at greater depths, using the well-optimized parameters obtained from the GRU-CNN as the starting point for the bilinear algorithm.

3.2 Loss Function

The choice of the loss function for the initial part of our network architecture (GRU) is determined by the enhancement achieved at each time step of the GRU unit, aggregated over the course of the optimization history. It is defined as follows:

$$L_{GRU}(\theta) = \sum_{t=1}^{T} \max\{E[GRU(\theta_t)] - \max_{i<t} E_{i<t}[GRU(\theta_i)], 0\} \quad (12)$$

where we calculate the discrepancy between the expectation at time step *t* and the highest expectation achieved during the prior history of optimization up to that moment. If no improvement occurs at a particular time step, it contributes nothing to the loss. A larger step size corresponds to a lower loss, signifying improved optimization effectiveness. However, it's important to note that increasing the step size leads to longer iteration times.

To train the GRU, it's necessary to compute the derivative of the aforementioned loss function, denoted as $L_{GRU}(\theta)$, and then apply the resulting gradient to the network. This process is commonly known as temporal backpropagation.

In the case of the CNN component, we utilize the average mean square error as our loss function, defined as follows:

$$L_{CNN}(\theta) = \frac{1}{n} \sum_{i=1}^{n} \|CNN(\theta_1) - \theta_2^*\|^2 \quad (13)$$

where n denotes the number of graphs in the training set, $\theta_1$ denotes the parameters obtained from the GRU optimization, and $\theta_2^*$ denotes the optimal parameters. The optimal parameter set for depth 2 QAOA, obtained using CNN, is determined by minimizing the loss function.

## 4. Performance of GRU-CNN on Max-Cut problem

4.1 Setup

We assess the efficacy of our proposed method by applying it to solve the Max-Cut problem on Erdős–Rényi graphs with varying node counts ($N \in [4, 14]$). In Erdős–Rényi graphs, the existence of each edge (*i, j*) is determined independently with a certain probability, irrespective of other edges. We examine two distinct graph ensembles:

(a) Random Erdős–Rényi graph nodes: In this configuration, we generate graph instances across a range of node counts ($N \in [4, 14]$) for a specified edge probability. These settings with different numbers of nodes allow us to evaluate the performance of the QAOA on varying qubit architectures.

(b) Random Erdős–Rényi graph edge probabilities: For this ensemble, the edge probabilities differ between graphs but remain uniform for all edges within a single graph. In this scenario, we independently sample edge probabilities within the range of $p \in [0.5, 1]$ for each graph and apply these probabilities to all edges. These settings with diverse edge probabilities offer a more realistic representation of real-world problem instances.

### 4.2 Implementation Details

To evaluate the performance of our approach, we scrutinized several well-established gradient-based optimization algorithms, such as Adam, which integrates momentum and time-dependent learning rates into its update mechanism, alongside RMSProp and Adagrad. To maintain uniformity, we utilized an identical random seed for the initialization procedure of each method, yielding consistent parameter values. Furthermore, we fixed the learning rate at 0.1. Through the comparison of our method's performance with these three initialization strategies, we can assess its efficacy. We utilized Erdős–Rényi graphs with node counts ranging from 4 to 14 to assess the performance of our method. We sampled the edges of these graphs with probabilities spanning from 0.5 to 1. For each parameter configuration, we generated 10 instances for testing and calculated their average approximation ratio for benchmarking. To optimize the QAOA algorithm, we applied distinct optimization techniques to each instance, including GRU, Adam, RMSProp, and Adagrad, to derive the Max-Cut solution. Note that all QAOA circuit executions were conducted using Qiskit's noise-free statevector simulator [39].

### 4.3 Experimental Results

In this subsection, we conducted numerical simulations to compare the performance of our approach with other optimization methods. The aim was to assess the efficacy of our network architecture in tackling the Max-Cut problem.

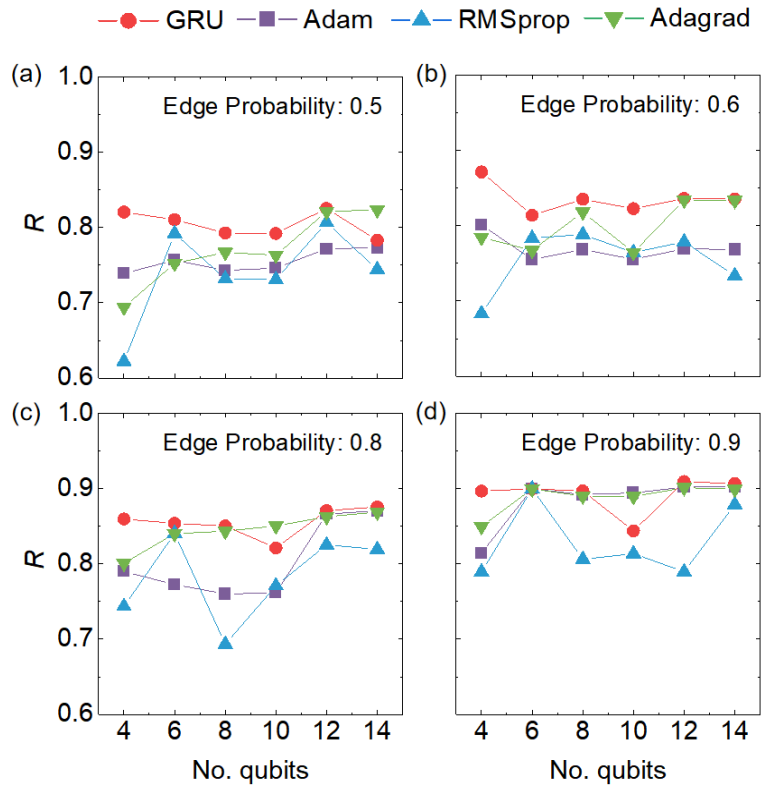

Figure 3: QAOA different initialization methods with different qubit sizes. The subplots in the figure illustrate the simulation results of QAOA for different edge probability settings, namely 0.5, 0.6, 0.8, and 0.9. From left to right, the subplots depict the simulation results of QAOA for the edge probability settings of 0.5, 0.6, 0.8, and 0.9. The x-axis represents the magnitude of the quantum bits, which grows from 4 to 14, the y-axis represents the approximation ratios corresponding to the tests for each setting. The four different curves correspond to the levels of the four initialization methods used for the QAOA circuits. We created 10 random problem instances for each configuration to conduct tests, and evaluated the performance based on the average approximation ratio attained by each method on these instances.

Fig. 3 illustrates various instances of Max-Cut graphs with different edge probabilities (0.5, 0.6, 0.8, and 0.9) for the depth 1 QAOA. In Fig. 3(a), we present the approximation ratio achieved by our optimization strategy in comparison to that of other classical optimizers for varying numbers of nodes (equivalent to the number of qubits) under the condition of 0.5 edge probability. It is evident that our proposed GRU stochastic initialization strategy consistently outperforms the other three classical optimizers (Adam, RMSProp, and Adagrad) in the majority of the tested instances. The approximation ratios attained by the GRU initialization method are notably higher and can be regarded as quasi-optimal initialization techniques. For instance, on graphs with 4 nodes and an edge probability of 0.5, the GRU initialization method attains an approximation ratio of 82%, representing a substantial improvement over the maximum approximation ratio of approximately 74% obtained by the other three optimizers.

In Figs. 3(b), 3(c), and 3(d), we further explore the approximation ratios achieved by the GRU optimization strategy as compared to the optimization strategies of the other classical methods for edge probabilities of 0.6, 0.8, and 0.9, respectively. The numerical findings consistently reveal that the GRU stochastic initialization method introduced in this paper yields higher approximation ratios than the classical optimizers in the majority of the test instances. This improvement is substantial and enhances the identification of favorable initial variational parameters.

Our attention now turns to a different scenario, where we specifically assess the approximation ratios achieved by each optimization strategy across varying edge probabilities within the range of $p \in [0.5, 1]$. In Fig. 4, we present the outcomes of our experiments, where we compare the approximation ratios accomplished by our strategy with those obtained by the other three gradient-based optimizers for node sizes of 6, 8, 12, and 14 as the edge probabilities increase. In Fig. 4(a), 4(b) and 4(c), corresponding to 6, 8, and 12 nodes, the GRU optimization strategy consistently surpasses the performance of the Adam, RMSProp, and Adagrad optimizers across edge probabilities ranging from 0.5 to 1. Notable enhancements in optimization performance are observed. In Fig. 4(d), the test results obtained by our strategy, comprising 10 instances for each setting with edge probabilities of 0.6, 0.8, 0.9, and 1 on 14 nodes, consistently outperform the optimization results achieved by the other three classical optimizers. Consequently, we have established a general QAOA random initialization method that outperforms classical optimizers on the vast majority of Max-Cut instances.

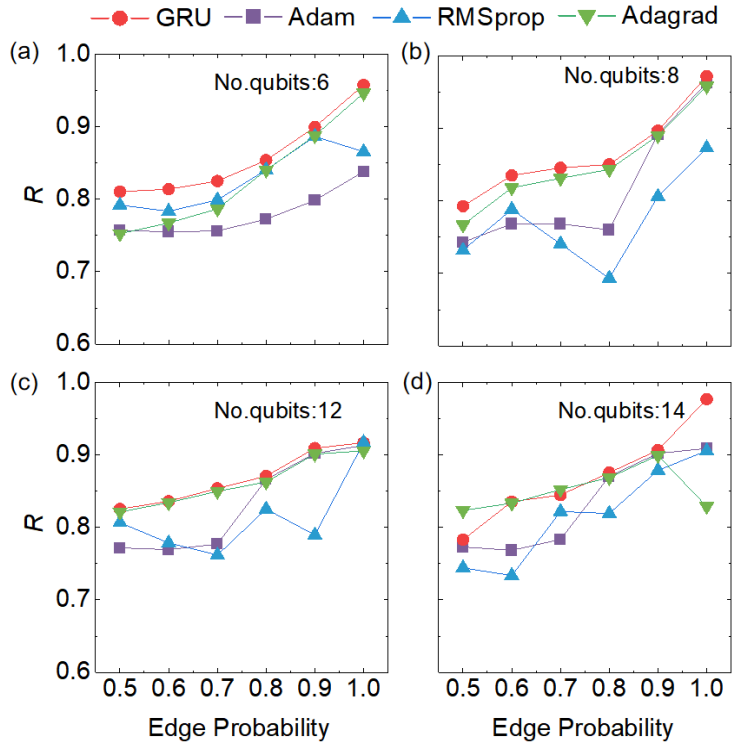

Figure 4: QAOA different initialization methods with different edge probability sizes. From left to right, the subplots depict the simulation results for QAOA with qubit settings of 6, 8, 12, and 14. The x-axis represents the magnitude of the edge probability, ranging from 0.5 to 1, while the y-axis represents the corresponding approximation ratios obtained from the tests for each setting. The four curves depict the levels of the four initialization methods used for the QAOA circuits. We created 10 random problem instances for each configuration to conduct tests, and evaluated the performance based on the average approximation ratio attained by each method on these instances.

For QAOA circuits with a depth of 2, we utilize a CNN as the foundational architecture for predicting the optimal parameters. The training of the CNN network involves the application of the Adam optimizer with an initial learning rate set at $1 \times 10^{-4}$. The training dataset comprises information regarding the optimal parameters for various graph configurations. We establish a batch size of 6 and conduct 50 epochs for training. The training progress of the CNN is illustrated in Fig. 5(a), demonstrating a gradual convergence of the loss to lower levels as the number of epochs increases. The use of well-initialized variational parameters, obtained through initialization by the GRU, leads to a robust initial prediction of QAOA circuit parameters. Subsequently, the CNN architecture is employed to enhance performance and leverage its capacity to extract parameter features, thereby facilitating effective parameter prediction for deeper circuit depths.

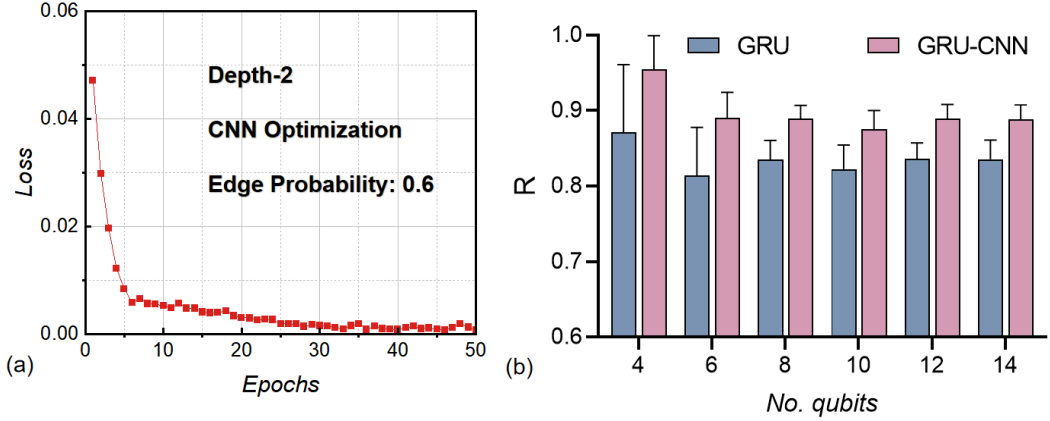

Figure 5: (a) Optimization of depth-2 circuits using CNNs with an edge probability of 0.6. The x-axis indicates the epoch number, and the y-axis signifies the corresponding training loss value. (b) QAOA performance compared to GRU and GRU-CNN for various qubit numbers, with R representing the approximation ratio. The values are averaged over 10 realizations for each node on the graph.

To provide a comprehensive evaluation of this approach, we undertake a systematic simulation involving graphs characterized by an edge probability of 0.6, as displayed in Fig. 5b. This simulation integrates CNN and GRU as QAOA (depth 2) initialization and optimization routines. Both methods are applied to randomly generated instances spanning from 4 to 14 nodes, with 10 instances generated for each configuration. Subsequently, we calculate the average approximation ratios to gauge the performance of QAOA. Our findings reveal that the GRU-CNN model significantly enhances the performance of QAOA compared to that of GRU initialization alone. This underscores the crucial role played by CNN in the optimization of QAOA.

## 5. Predicting Higher Depth QAOA Parameters Using Bilinear Algorithm

In this section, we apply our GRU-CNN initialization optimization strategy to the bilinear approach [14] and conduct a comprehensive analysis of the performance of QAOA at higher depths on graph instances.

Previous studies[14, 36, 40-42] have indicated that the optimal parameter values for QAOA instances exhibit non-random patterns and certain regularities [37, 43]. Specifically, at a given index $j$, the optimal values of the mixing layer parameters ($\beta_l^*$) for any QAOA instance with depth-$l$ gradually decrease between steps (or stages), while the optimal values of the phase separating parameters ($\gamma_l^*$) increase between steps. As the depth $l$ increases, the optimal parameters of QAOA tend to follow a pattern that approximates linearity. Note that the pattern is not completely linear, and it is expected to approach linearity as $l$ approaches infinity($l\rightarrow\infty$) . Additionally, at the same depth $l$, $\gamma_j^*$ increases linearly with $j$ and $\beta_j^*$ decreases linearly with $j$. Based on the observed patterns, we can utilize the difference between the optimal parameters of the first two depths to predict the optimal initial variational parameters for higher depths using linear interpolation. his approach ensures that the initial parameters generated are in proximity to the optimal parameters, reducing the likelihood of converging to an undesired optimum. Fig. 6(a) provides an illustration of this strategy. To implement it, we begin by employing GRU-CNN to determine the optimal solutions for $l=1$ and $l=2$, forming the foundation for our strategy. We then calculate the difference between these two sets of optimal parameters to capture the pattern. Subsequently, we apply the strategy, commencing from $l=3$. The prediction method is outlined as follows:

$$\theta_l^j = \theta_{l-1}^j + (\theta_{l-1}^j - \theta_{l-2}^j), \quad j \leq l-2 \tag{14}$$

For the parameters with an index $j=l-1$, given that parameters $\theta_{l-1}^{l-2}$ do not exist, we use the previous index $j=l-2$ for prediction:

$$\theta_l^{l-1} = \theta_{l-1}^{l-1} + (\theta_{l-2}^{l-1} - \theta_{l-2}^{l-2}) \tag{15}$$

For the new parameter $j=l$, update as follows:

$$\theta_l^l = \theta_l^{l-1} + (\theta_l^{l-1} - \theta_l^{l-2}) \tag{16}$$

We apply the bilinear strategy in conjunction with our GRU-CNN preprocessing approach to tackle the maximum cut problem on Erdős-Rényi graphs using higher-depth QAOA. The effectiveness of our strategy is evaluated across various instances of graphs with node numbers of 8, 10, and 12, and an edge probability of 0.5. In Fig. 6(b), we present the approximation ratios for QAOA circuit depths ranging from 1 to 12. For depths 1 and 2, the QAOA circuit parameters are predicted by our GRU-CNN model, resulting in their respective approximation ratios. Leveraging the quasi-optimal parameters obtained for the initial two depths, we extend the parameters of the quantum circuits to higher depths using the bilinear strategy. Notably, for qubit numbers 8, 10, and 12, our strategy achieves approximation ratios of 0.995, 0.998, and 0.996, respectively, for the QAOA circuit at depth 12. Our strategy attains an average approximation ratio close to the optimum, underscoring its effectiveness. It's important to note that the observed trends apply to all other instances as well.

We conducted a comparison between our bilinear algorithm and two other machine learning-based strategies, namely random initialization and the PPN$^2$-based strategy [15]. This comparison was performed on a dataset comprising 8-node graphs, each with an edge probability of 0.6. As shown in Fig. 6(c), our proposed bilinear strategy achieves an approximation ratio of up to 0.9874. Compared to random initialization, the average approximation ratio is 9.4% higher. Random initial values often constrain the QAOA parameters, resulting in a high number of function calls. In addition, the PPN$^2$-based strategy predicts parameters for the next depth by seeking the parameter with the highest expected value in the parameter set. However, due to the constraints of the training data, the bilinear strategy introduced in this paper establishes a higher approximation upper bound compared to PPN$^2$. This enables us to predict parameters for higher-depth QAOA. The bilinear strategy achieved a prediction of QAOA at a depth of 10 and produced better results than the other two machine learning based strategies on our test set. Meanwhile, compared to existing machine learning based strategies that require a large amount of training data to achieve competitive results, our strategy greatly reduces this demand.

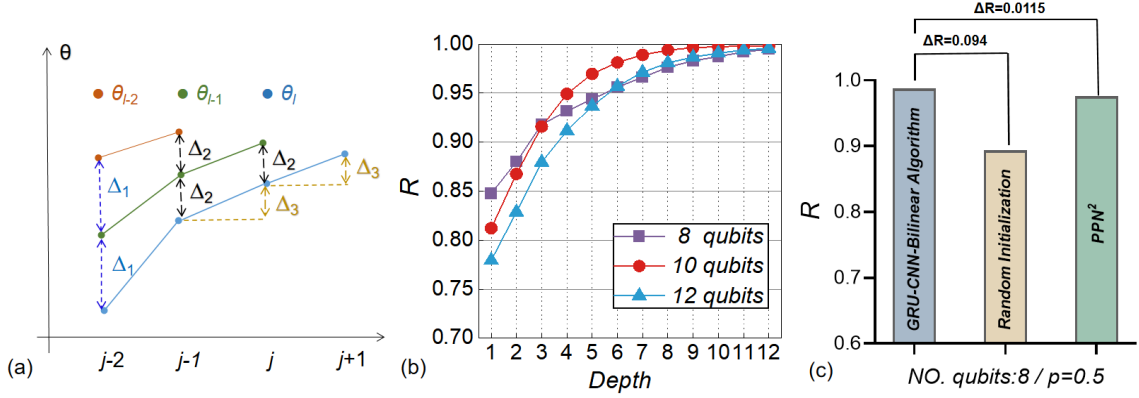

Figure 6: (a) Schematic diagram of bilinear algorithm. (b) Approximation ratio achieved on QAOA at depths of 1-12 using GRU-CNN preprocessing combined with bilinear strategy at 8, 10, and 12 qubits. Each setting is implemented more than 10 times on average. (c) The average results of approximate ratios of bilinear strategy, random initialization, and $PPN^2$-based strategy on the test set.

## 6. Conclusion and outlook

In this study, we introduce a hybrid GRU-CNN neural network model for predicting initial variational parameters ($\gamma$, $\beta$) in depth 1 and 2 QAOA circuits, employed to solve the Max-Cut problem for each graph instance. Our approach outperforms classical gradient-based optimizers in terms of approximation ratios for solving the Max-Cut problem on the Erdős-Rényi graph. Building upon this, we utilize a hybrid GRU-CNN neural network optimization strategy to preprocess QAOA circuits with depths of 1 and 2, generating corresponding initial parameters, which are then applied to the bilinear algorithm for higher circuit depths. This preprocessing method can effectively generate good initial parameters ($l$=1,2) for different Max-Cut problem instances. As a result, we successfully predict depth-12 QAOA circuit parameters for 8, 10, and 12 qubits, with results closely approaching optimality. In further benchmarking against other machine learning-based strategies, our strategy exhibits significant performance advantages, demonstrating that it is not constrained by training data and significantly improving the algorithm's generalization performance.

With the advancement of quantum computers and their growing number of qubits, the ability to address larger instances of computational problems becomes feasible. This progress creates new opportunities for solving complex problems that were previously intractable for classical computers. Quantum computation empowers researchers to develop innovative solutions to real-world challenges more effectively.

However, for practical solutions, it is essential to limit the number of layers in QAOA, especially in the NISQ era, where the increased size of quantum devices introduces significant noise for executing deep circuits [28]. Our personalized prediction of the initial variational parameters of QAOA circuits using hybrid neural networks becomes particularly practical in this context. Through benchmarking, we have demonstrated that our method outperforms classical optimizers in terms of approximation ratios. This ongoing research direction holds promise for further advancements in quantum computing applications.

## Acknowledgments

This work was partly supported by the National Natural Science Foundation of China (Grant Nos. 61874001, 62004001, 62201005, 62004001, 62304001), the Anhui Provincial Natural Science Foundation under Grant No. 2308085QF213, and the Natural Science Research Project of Anhui Educational Committee under Grant No. 2023AH050072.

## References

1. Cerezo, M., et al., *Variational quantum algorithms.* Nature Reviews Physics, 2021. **3**(9): p. 625-644.
2. Preskill, J., *Quantum Computing in the NISQ era and beyond.* Quantum, 2018. **2**: p. 20.
3. Zhou, L., et al., *Quantum Approximate Optimization Algorithm: Performance, Mechanism, and Implementation on Near-Term Devices.* Physical Review X, 2020. **10**(2): p. 23.
4. Farhi, E., J. Goldstone, and S. Gutmann, *A Quantum Approximate Optimization Algorithm.* arXiv pre-print server, 2014.
5. Ruan, Y., et al., *The Quantum Approximate Algorithm for Solving Traveling Salesman Problem.* Computers, Materials \& Continua, 2020. **63**(3): p. 1237--1247.
6. Andrew, L., *Ising formulations of many NP problems.* Frontiers in Physics, 2013. **2**: p. 5.
7. Kochenberger, G.A., et al., *The unconstrained binary quadratic programming problem: a survey.* Journal of Combinatorial Optimization, 2014. **28**: p. 58-81.
8. Barkoutsos, P.K., et al., *Improving Variational Quantum Optimization using CVaR.* Quantum, 2020. **4**: p. 256.
9. Guerreschi, G.G. and A.Y. Matsuura, *QAOA for Max-Cut requires hundreds of qubits for quantum speed-up.* Scientific Reports.
10. Niu, M.Y., S. Lu, and I.L. Chuang, *Optimizing QAOA: Success Probability and Runtime Dependence on Circuit Depth.* 2019.
11. Jain, N., et al., *Graph neural network initialisation of quantum approximate optimisation.* Quantum, 2022. **6**: p. 25.
12. Wierichs, D., C. Gogolin, and M. Kastoryano, *Avoiding local minima in variational quantum eigensolvers with the natural gradient optimizer.* 2020.
13. Verdon, G., et al., *Learning to learn with quantum neural networks via classical neural networks.* arXiv preprint arXiv:1907.05415, 2019.

14. Lee, X., et al., *A Depth-Progressive Initialization Strategy for Quantum Approximate Optimization Algorithm.* Mathematics, 2023. **11**(9): p. 2176.

15. Xie, N., et al., *Quantum Approximate Optimization Algorithm Parameter Prediction Using a Convolutional Neural Network.* arXiv preprint arXiv:2211.09513, 2022.

16. Verdon, G., et al., *A Quantum Approximate Optimization Algorithm for continuous problems.* arXiv pre-print server, 2019.

17. Amaro, D., et al., *Filtering variational quantum algorithms for combinatorial optimization.* Quantum Science and Technology, 2022. **7**(1): p. 17.

18. Peruzzo, A., et al., *A variational eigenvalue solver on a photonic quantum processor.* Nature Publishing Group, 2014(1).

19. Killoran, N., et al., *Continuous-variable quantum neural networks*. 2018.

20. Troyer, et al., *Progress towards practical quantum variational algorithms.* Physical Review A Atomic Molecular & Optical Physics, 2015.

21. Lilienfeld, A.V. *Quantum Machine Learning*. in *APS March Meeting 2017*. 2017.

22. Bharti, K., et al., *Noisy intermediate-scale quantum (NISQ) algorithms.* 2021.

23. Diederik and J. Ba, *Adam: A Method for Stochastic Optimization.* arXiv pre-print server, 2017.

24. Lydia, A. and S. Francis, *Adagrad - An Optimizer for Stochastic Gradient Descent.* 2019.

25. Igor, et al., *Simulating Quantum Computation by Contracting Tensor Networks.* SIAM Journal on Computing, 2008. **38**(3): p. 963–981.

26. Wang, J., et al. *Analysis of influence factors in Quantum Approximate Optimization Algorithm for Solving Max-cut Problem*. IEEE.

27. Pranav Chandarana, P.S.V., Narendra N. Hegade, Enrique Solano, Yue Ban, Xi Chen, *Meta-learning digitized-counterdiabatic quantum optimization.* arXiv pre-print server, 2022.

28. Amosy, O., et al., *Iterative-free quantum approximate optimization algorithm using neural networks.* arXiv preprint arXiv:2208.09888, 2022.

29. Majumdar, R., et al., *Optimizing Ansatz Design in QAOA for Max-cut.* arXiv pre-print server, 2021.

30. Wong, R.T., *Combinatorial Optimization: Algorithms and Complexity (Christos H. Papadimitriou and Kenneth Steiglitz).* SIAM Review, 1983. **25**(3): p. 424-2.

31. Cho, K., et al., *On the Properties of Neural Machine Translation: Encoder-Decoder Approaches.* Computer Science, 2014.

32. Lechner, M. and R. Hasani, *Learning Long-Term Dependencies in Irregularly-Sampled Time Series.* 2020.

33. Noh, S.-H., *Analysis of Gradient Vanishing of RNNs and Performance Comparison.* Information, 2021. **12**(11): p. 442.

34. Yongkang, et al., *A novel attention-based hybrid CNN-RNN architecture for sEMG-based gesture recognition.* PloS one, 2018.

35. Technicolor, T.S., et al., *ImageNet Classification with Deep Convolutional Neural Networks [50]*.

36. Alam, M., A. Ash-Saki, and S. Ghosh. *Accelerating Quantum Approximate Optimization Algorithm using Machine Learning*. IEEE.

37. Crooks, G.E., *Performance of the Quantum Approximate Optimization Algorithm on the Maximum Cut Problem.* 2018.

38. Abien, *Deep Learning using Rectified Linear Units (ReLU)*. arXiv pre-print server, 2019.
39. Cross, A. *The IBM Q experience and QISKit open-source quantum computing software*. in *APS March Meeting 2018*. 2018.
40. Sack, S.H. and M. Serbyn, *Quantum annealing initialization of the quantum approximate optimization algorithm*. 2021.
41. Willsch, M., et al., *Benchmarking the Quantum Approximate Optimization Algorithm*. 2019.
42. Cook, J., S. Eidenbenz, and A. Brtschi. *The Quantum Alternating Operator Ansatz on Max-k Vertex Cover*. in *APS March Meeting 2020*. 2020.
43. Zhou, L., et al., *Quantum Approximate Optimization Algorithm: Performance, Mechanism, and Implementation on Near-Term Devices*. 2018.